\documentclass[final,times,5p,twocolumn]{elsarticle}
\biboptions{sort&compress}
\usepackage{amsmath,amssymb}
\usepackage{color}
\usepackage{graphicx}
\usepackage{tabularx}
\usepackage{lineno}\modulolinenumbers[5]
\usepackage[euler]{textgreek}
\usepackage{siunitx}
\DeclareSIUnit{\wtpercent}{wt.\%} %For siunitx
\DeclareSIUnit{\atpercent}{at.\%} %For siunitx
\usepackage{miller}
\usepackage{hyperref}

\journal{Nature Communications}

\begin{document}

\begin{frontmatter}

\title{\textcolor{black}{The influence of alloying on slip intermittency and the implications for dwell fatigue in titanium}} 

\author[IC]{Felicity F. Worsnop\corref{cor1}}
\author[PS]{Rachel E. Lim}
\author[LLNL]{Joel V. Bernier}
\author[PS]{Darren C. Pagan}
\author[IC]{Yilun Xu}
\author[IC]{Thomas P. McAuliffe}
\author[RR]{David Rugg}
\author[IC]{David Dye\corref{cor1}}

\address[IC]{Department of Materials, Royal School of Mines, Imperial College London, Prince Consort Road, London, SW7 2AZ, UK}
\address[PS]{Department of Materials Science and Engineering, Pennsylvania State University, 221 Steidle Building, University Park, PA 16802}
\address[LLNL]{Lawrence Livermore National Laboratory, 7000 East Avenue, Livermore, CA 94550, USA}
\address[RR]{Formerly with Rolls-Royce plc., Elton Road, Derby, DE24 8BJ, UK}
\cortext[cor1]{Corresponding author}

\begin{abstract}
Dwell fatigue, the reduction in fatigue life experienced by titanium alloys due to holds at stresses as low as 60\% of yield, has been implicated in several uncontained jet engine failures. Dislocation slip has long been observed to be an intermittent, scale-bridging phenomenon, similar to that seen in earthquakes but at the nanoscale, leading to the speculation that large stress bursts might promote the initial opening of a crack. Here we observe such stress bursts at the scale of individual grains \textit{in situ}, using high energy X-ray diffraction microscopy in Ti--7Al--O alloys. This shows that the detrimental effect of precipitation of ordered Ti$_3$Al is to increase the magnitude of rare pri\hkl<a> and bas\hkl<a> slip bursts associated with slip localisation. By contrast, the addition of trace O interstitials is beneficial, reducing the magnitude of slip bursts and promoting a higher frequency of smaller events. This is further evidence that the formation of long paths for easy basal plane slip localisation should be avoided when engineering titanium alloys against dwell fatigue.\end{abstract}

\begin{keyword}
Titanium alloys \sep Plasticity \sep Slip avalanche
\end{keyword}

\end{frontmatter}
%\linenumbers

\section{Introduction}
\label{intro}

Since the earliest \textit{in situ} TEM observations of dislocations, dislocation motion has been observed to be intermittent, with long waiting times in between bursts of activity, more recently termed dislocation avalanches \cite{Miguel2001,Weiss2007,Brown2012,Sparks2018}. By contrast, most macroscopic phenomenological models of time-dependent plastic deformation are viscoplastic, spring-and-dashpot models after Zener \cite{Christensen1982}. Crystal plasticity finite element models of assemblages of single crystals into microstructures typically also follow this approach, but the discrete dislocation plasticity approach allows such intermittent motion of individual dislocations to be modelled, enabling more realistic treatment of phenomena such as the interaction of slip bands with grain boundaries \cite{Zheng2016}. However, the implications of dislocation avalanches for material performance in the field have yet to be elucidated. To date, it has been an interesting scientific phenomenon without obvious application.

Cold dwell fatigue is a phenomenon in hexagonal \textalpha{}-titanium where holds at elevated stress give rise to reductions in cyclic lifetime of more than an order of magnitude, even at stresses substantially below the yield stress \cite{Bridier2008,Neal1988,Dunne2007,AFRL1757}. The emerging understanding of dwell fatigue is that it is a result of \hkl<a> slip bands active in `soft' orientations running across long easy slip paths in regions of common orientation inherited from processing, termed macrozones, which then impinge on `hard' oriented grains poorly oriented for slip, in turn resulting in stress localisation that nucleates cracking \cite{Dunne2007}. Ageing of \textalpha{}$_2$ Ti$_3$Al, which is a superstructure of hexagonal \textalpha{}-Ti, is therefore detrimental because \hkl<a> dislocations are then required to travel in pairs which suppresses cross-slip and promotes slip band formation \cite{Neeraj2000}. For this reason high Al$_\mathrm{eq}$ contents are also detrimental, which is why dwell fatigue was first observed in alloys such as Ti-811 and IMI685 \cite{Bache2003}. However, the role of oxygen in dwell fatigue is less well established; O is an interstitial solid solution strengthener in Ti alloys but beyond around 2500~ppmw O additions are detrimental to ductility and fatigue life. However, at lower O contents recent \textit{in situ} TEM observations and DFT modelling have suggested that O atoms are pushed out of the the path of moving dislocations \cite{Yu2015}, creating obstacle-free channels, but also that O additions result in very viscous, heavily pinned dislocation motion \cite{Barkia2017,Chong2020}.
\begin{figure*}[t!]
   \centering
   \includegraphics[width=178mm]{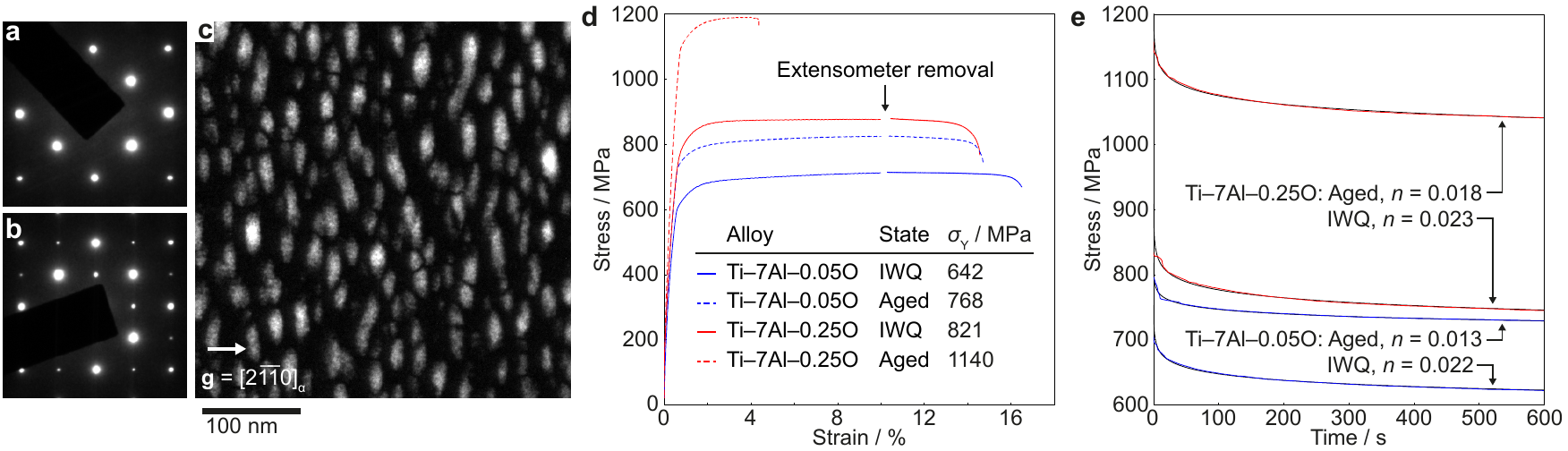}
   \caption{Selected area electron diffraction patterns (\textbf{B} = \hkl[0 1 -1 1]) for the \textbf{a} quenched and \textbf{b} aged conditions show the appearance of superlattice reflections from the \textalpha{}$_2$ precipitates, shown in \textbf{c} for Ti--7Al--0.25O (dark field, \textbf{B} = \hkl[0 1 -1 1], \textbf{g} = \hkl[2 -1 -1 0] two-beam condition). Macroscopic mechanical responses and microstructure of Ti--7Al--O alloys. \textbf{d} Ageing to produce crystallographically ordered \textalpha{}$_2$ Ti$_3$Al precipitates causes strengthening but with loss of ductility. Raising the interstitial O content also provides a strengthening effect. \textbf{e} Stress relaxation testing revealed that \textalpha{}$_2$ affected the rate constant to a greater extent than interstitial O.}
   \label{fig:micro-stressrel}
\end{figure*}

High energy X-ray diffraction microscopy (HEDM) is a technique developed over the last decade that allows the back-projection and deconvolution of diffraction spots from individual grains to reconstruct the location and elastic strain state of lightly deformed O(\SI{50}{\micro\metre}) grains in O(\SI{1}{\milli\metre\cubed}) volumes of primarily metallic materials. Leading applications have been combined with near-field reconstructions of grain shapes in a tomographic-type approach, to the study of recrystallisation and of the grain-average stress tensor in deforming materials \cite{Bernier2011,Pagan2017}. The technique generates large volumes of data, which itself has represented a significant obstacle to application. A particular problem has been the application to fatigue and crack initiation, given the small volumes studied compared to the stressed volume in engineering components. For example, the stressed volume of critically stressed regions in even a small jet engine fleet of fan discs  might be on the order of \SI{E9}{\milli\metre\cubed}, meaning that the critical microstructural feature causing an unacceptable rate of service failure (\emph{i.e.} ever) might be quite unlikely to be contained within any test sample. Recently, we have attempted to overcome this by studying the hypothesized feature of interest, intentionally introduced into the microstructure \cite{XuJoseph2020}. \textcolor{black}{Experimentally, micropillar compression and acoustic emission spectroscopy have been used to study intragranular slip intermittency \cite{Ispanovity2022}. To elucidate the implications of dislocation avalanches in engineering contexts, a mesoscopic approach able to capture the behaviour of real microstructures is needed.} In previous work, Beaudoin et al. found that the stress drops associated with dislocation avalanches \textcolor{black}{could} be observed by HEDM in Ti--7Al \textalpha{}-Ti, and that the largest stress drops observed were associated with hard/soft grain combinations, as expected \cite{Beaudoin2017}.

Here we turn to address the question of Al content (\textalpha{}$_2$ Ti$_3$Al formation) and interstitial O and their role in dwell fatigue, critical questions for the alloy designer, by examination of their effect on such stress drops within microstructures (rather than micropillars) using HEDM.

\section{Results}
\label{results}

Ti--7Al alloys containing 0.05 and 0.25 \si{\wtpercent} oxygen were tested, providing low and high oxygen contents relevant to engineering applications. Specimens were processed to generate equiaxed \textalpha{} microstructures with grain sizes of 50--\SI{100}{\micro\metre}, Fig.~\ref{fig:supplementary-sem}, and were then heat treated either to induce crystallographic disorder of Al on the hcp \textalpha{} lattice or to produce a homogeneous dispersion of \textalpha{}$_2$ Ti$_3$Al precipitates, Fig.~\ref{fig:micro-stressrel}\textcolor{black}{(a--c)}. In these four conditions, tensile testing was conducted at a strain rate of \SI{E-3}{\per\second} and showed the significant solid solution strengthening effect of interstitial O, and the precipitation hardening effect of \textalpha{}$_2$, Fig.~\ref{fig:micro-stressrel}\textcolor{black}{(b)}. \textcolor{black}{Raised oxygen content promotes a finer \textalpha{}$_2$ dispersion, quantified in our previous study \cite{Dear2021}, reflected by the significantly higher yield stress for the high-oxygen alloy than the low-oxygen case after ageing. The observed loss of ductility at high oxygen contents is very commonly seen \cite{Lim1976}, particularly above 2300 ppmw.}

\textbf{Stress relaxation tests.} Stress relaxation behaviour, relevant to time-dependent damage mechanisms such as dwell fatigue, was characterised for each material by mechanical testing followed by \textit{post mortem} TEM imaging of dislocations. Strain-controlled loading was carried out on specimens with a 3$\times$\SI{3}{\milli\metre} gauge section at a strain rate of \SI{E-4}{\per\second} during initial loading, achieving 0.5--0.8\% plastic strain, followed by a \SI{600}{\second} hold in strain control. The responses of each material are shown in Fig.~\ref{fig:micro-stressrel}(e), and were fitted with power law decay curves of the form $\sigma = Ct^{-n}$ to enable comparison of decay exponents $n$ (where $C$ is a material constant). Results indicated that the decay constant was more dependent on ageing condition than on O content, indicating that the locations of Al atoms on the hcp lattice have a rate-controlling effect on slip during relaxation of stresses after yielding, while interstitial O appears less influential.

TEM specimens extracted from the central gauge region of each sample were then characterised to observe the resulting dislocation configurations. It has long been observed that both Al and O can cause restriction of dislocation motion to narrow slip bands \cite{Williams1972,Neeraj2001}. In dilute Ti--Al alloys, pairing of dislocations within a slip band is also seen where crystallographic ordering of Al has been induced during thermomechanical processing \cite{Neeraj2000}.

\begin{figure}[t!]
   \centering
   \includegraphics[width=88mm]{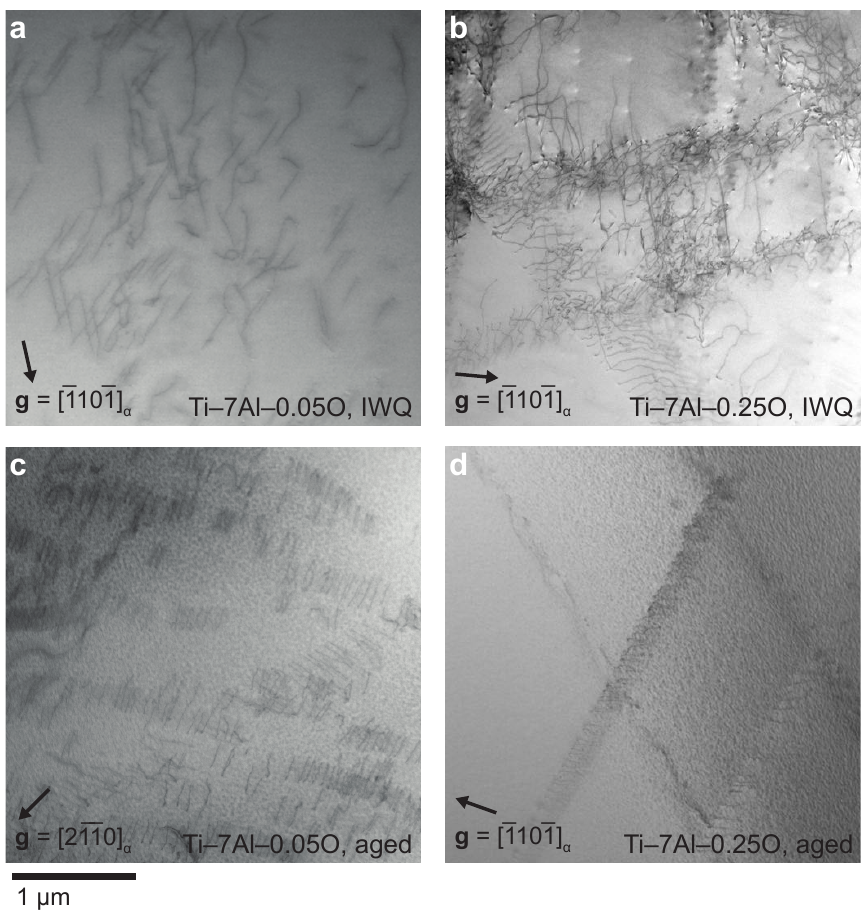}
   \caption{Examples of dislocation configurations after stress relaxation in Ti--7Al--0.05O and Ti--7Al--0.25O, in IWQ and aged conditions (\SI{550}{\celsius} for \SI{49}{\day}). STEM micrographs recorded for two-beam condition around \textbf{B} = \hkl{0 1 -1 1} as indicated. \textbf{a} In the low-oxygen, disordered material, dislocations are able to travel freely across a grain. Some dislocation pairing is evident, likely due to short-range order of aluminium that was not suppressed in quenching at such high Al content. \textbf{b} The addition of 2000~ppmw~O causes slip band formation, but also promotes significant cross-slip and resulting dislocation interactions. \textbf{c} Ageing to precipitate ordered \textalpha{}$_2$ causes more restricted slip bands, with less cross-slip possible due to the formation of antiphase boundaries between leading and trailing dislocations passing through ordered precipitates. \textbf{d} In the high-oxygen alloy, the \textalpha{}$_2$ dispersion is refined \cite{Dear2021}, causing even more severe slip banding. \textcolor{black}{The mottled contrast in \textbf{c} and \textbf{d} is from the \textalpha{}$_2$ precipitates.}}
   \label{fig:tem-dislocations}
\end{figure}

As expected, the quenched Ti--7Al--0.05O material showed the least spatially restricted dislocation configurations, Fig.~\ref{fig:tem-dislocations}(a). There is some evidence of incipient slip band formation, and instances of dislocations travelling in pairs. This suggests that a small degree of short-range order was present that could not be fully suppressed during quenching from solution temperatures due to the high Al content. With increased interstitial O content, slip band formation is promoted, but with significant activity outside the slip bands and abundant evidence of dislocation pinning and bowing, Fig.~\ref{fig:tem-dislocations}(b).  There is also minimal evidence of dislocation pairing in this case, as expected in the absence of crystallographic ordering of Al.

The slip bands caused by \textalpha{}$_2$ ordered precipitates are qualitatively different to those induced by raised oxygen content. Slip bands, paired dislocations and pinning by precipitates were observed for both the aged materials, with very little activity outside the bands, Fig.~\ref{fig:tem-dislocations}(c--d). In the aged high-oyxgen alloy, the restriction of dislocations to slip bands is even more severe. We suggest this is due to the higher strength of the surrounding unslipped material, due to refinement of \textalpha{}$_2$ by raised O content \cite{Dear2021}. This increases the difficulty of setting up new slip bands, and of dislocation motion outside bands. Considering also the macroscopic yield stress differences, classical Orowan bowing versus shearing considerations indicate that precipitate shearing is in effect. This is consistent with independent observations of shearing of ordered domains \cite{Zhang2019}.

\textbf{\textit{In situ} creep and ff-HEDM.} An \textit{in situ} creep ff-HEDM experiment was performed at the ID3A beamline at CHESS to probe the intermittency of slip in the quenched and aged states of the Ti--7Al--0.05O and Ti--7Al--0.25O alloys.

\begin{figure}[t!]
   \centering
   \includegraphics[width=88mm]{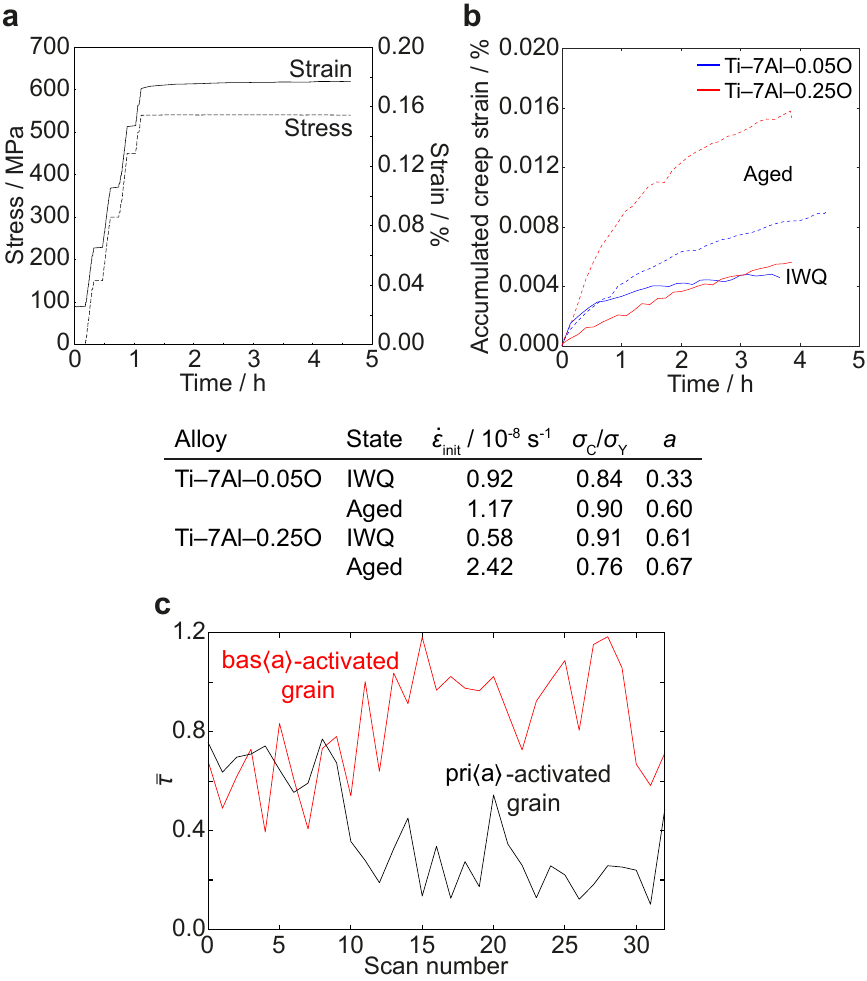}
   \caption{\textit{In situ} creep experiments with ff-HEDM allowed investigation of time-dependent micromechanics. \textbf{a} Example of macroscopic loading conditions, with steps in applied stress to achieve the desired initial strain rate. \textbf{b} Creep strain response of low- and high-oxygen alloys in the quenched and aged conditions. The table shows the creep stress for each sample as a fraction of its yield stress ($\sigma_{\mathrm{C}}/\sigma_{\mathrm{Y}}$), initial creep strain rates $\dot{\epsilon}_{\mathrm{init}}$ (averaged over the first hour) and the creep strain exponent \textit{a} obtained by fitting a power law relation. \textbf{c} The ff-HEDM data allowed tracking of individual grains' behaviour over time. Shown here for two neighbouring grains with different operative slip system types, the normalised resolved shear stress $\bar{\tau}$ varies over time, with evidence of load shedding in aged Ti--7Al--0.05O.}
   \label{fig:creep-graindrops}
\end{figure}
\begin{figure*}[t!]
   \centering
   \includegraphics[width=178mm]{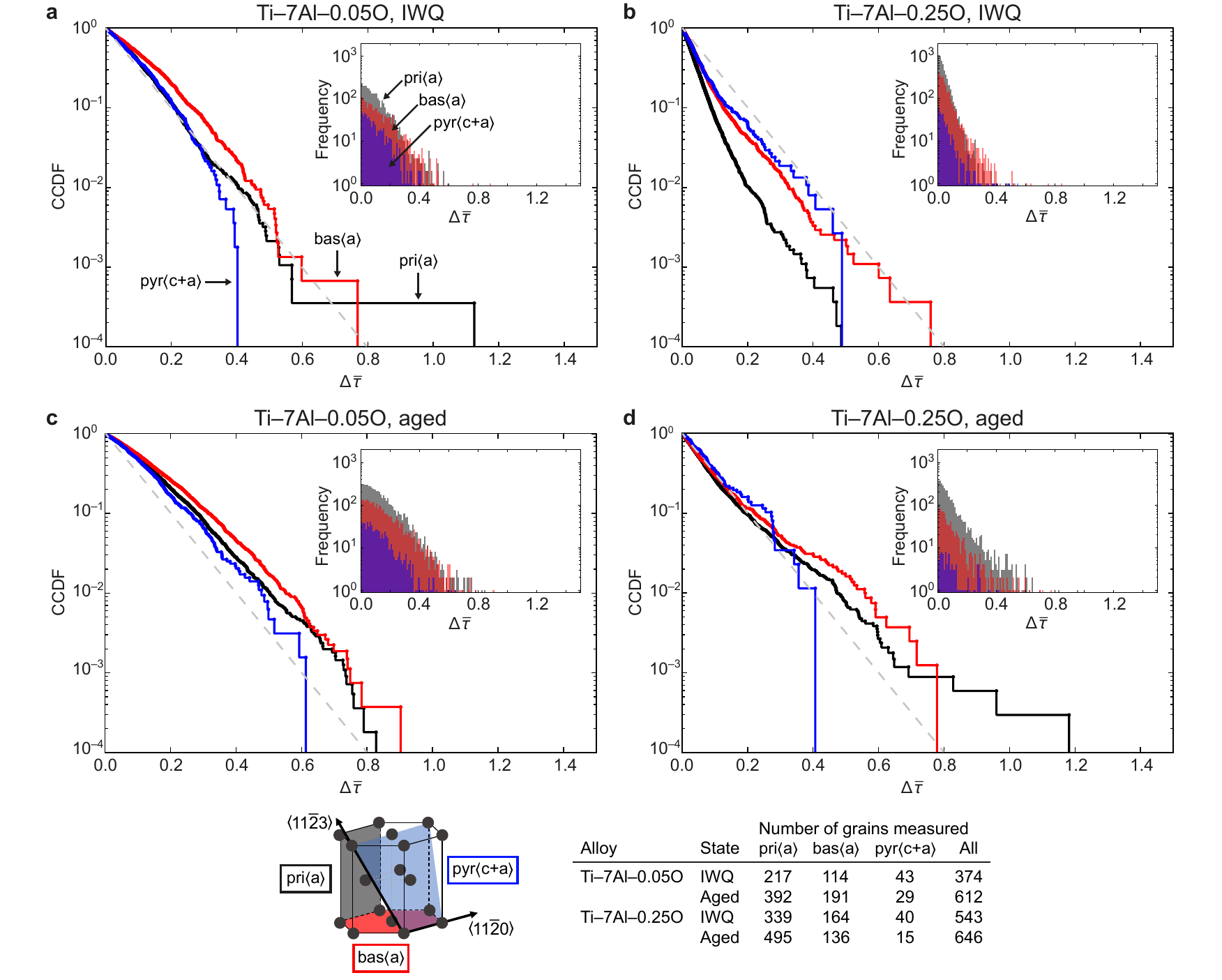}
   \caption{Cumulative complementary distribution functions (CCDFs) and histograms (inset) show the distribution of stress drop event sizes, $\Delta{}\bar{\tau}$. \textcolor{black}{In each case, several thousand stress drop events are captured, in a ratio of approximately 2.5:1:0.1 prismatic:basal:pyramidal.} The grey dashed lines are a guide to the eye, and the schematic unit cell provides a key for the slip systems shown.  In each case, bas\hkl<a> events are coarser than pri\hkl<a> events. \textbf{a} In disordered Ti--7Al--0.05O, there is a distribution of several small events, with a tail of larger drops. \textbf{b} Increasing interstitial O content shifts the distribution to smaller stress drops, corresponding with repeated cross-slip and pinning/unpinning events. \textbf{c} Ageing to induce formation of \textalpha{}$_2$ ordered domains causes an overall coarsening of slip events, \textbf{d} more pronounced for the high-oxygen alloy with its refined precipitate dispersion.}
   \label{fig:results-hedmstats}
\end{figure*}
Creep conditions were selected to account for the differing yield stresses of each material, aiming for a similar initial creep strain rate in each specimen. This was achieved at similar fractions of yield stress ($\sigma_{\mathrm{C}}/\sigma_{\mathrm{Y}}$) in each material except the aged high-oxygen alloy, which began to creep at a lower $\sigma_{\mathrm{C}}/\sigma_{\mathrm{Y}}$ than the other alloys (though a higher absolute stress). Creep strain--time curves for each hold, Fig.~\ref{fig:creep-graindrops}(b), were fitted using the power law creep relation, $\epsilon{} = At^{a}$ (creep time exponent $a$, load- and material-dependent prefactor $A$). Smaller values of $a$ correspond to more rapid exhaustion of primary creep, that is, the material proceeds towards steady-state creep more readily. Ti alloys undergo unusual amounts of primary creep and show significantly higher $a$ than other engineering alloys at room temperature \cite{Neeraj2000}. Here, $a$ was found to be higher in the presence of both higher O content and \textalpha{}$_2$ precipitates, indicating that both of the microstructural features that promote slip band formation also prolong primary creep.

During creep loading, ff-HEDM scans were run continuously at intervals of approximately \SI{6}{\minute} throughout the 3--\SI{4}{\hour} holds\textcolor{black}{, Fig.~\ref{fig:creep-graindrops}(b)}. For each grain, the centroid position, orientation and elastic strain tensor were obtained at each time interval during the creep hold. The resolved shear stresses $\tau$ were then calculated for each of the possible bas\hkl<a>, pri\hkl<a> and pyr\hkl<c+a> slip systems. For comparison between slip systems, these values were normalised against the critical resolved shear stress, $\tau^{*}$, for each system. \textcolor{black}{CRSS values used were 350, 330 and \SI{490}{\mega\pascal} for the prismatic \hkl<a>, basal \hkl<a> and pyramidal \hkl<c+a> slip system families, respectively, after Warwick et al. \cite{Warwick2012}.} The system with the highest $\bar{\tau} = \tau/\tau^{*}$ at the onset of creep was taken to be the active system during deformation, and the evolution of stresses on this system for each grain was then tracked to compare slip intermittency.

From one scan to the next, $\bar{\tau}$ may either rise or fall, as shown for two neighbouring grains in Fig.~\ref{fig:creep-graindrops}(c). A rise is caused by redistribution of micromechanical loads onto the grain of interest, due to deformation of neighbouring grains. Stress drops are the result of slip events occuring within the grain of interest, and it is these drops, $\Delta{}\bar{\tau}$, that are analysed here to assess the smoothness or jerkiness of slip in each of the four materials. It is noted that, during each of the \SI{6}{\minute} scan intervals, the ff-HEDM measurements may average over multiple dislocation motion events rather than resolving them individually.

After calculating all the stress drops on the active slip system for each grain throughout the creep hold, a set of complementary cumulative distribution functions (CCDFs) was plotted for each alloy, Fig.~\ref{fig:results-hedmstats}. Normalised against the numbers of grains and time intervals  for each sample, these allow direct comparison of the shapes of the distributions. The relative positions of curves to the left or right of the plot show which materials and slip systems experience larger or smaller events overall.

Comparing different slip systems in the quenched low-oxygen alloy, Fig.~\ref{fig:results-hedmstats}(a), we see that the distribution of event sizes for bas\hkl<a> slip lies at larger values than for pri\hkl<a>. This behaviour is common to all four materials studied, and indicates that basal slip proceeds by larger but fewer avalanches, with coarser intermittency.

Comparing the low-O and high-O alloys in the quenched condition, Fig.~\ref{fig:results-hedmstats}(a) and (b), there is a significant shift in behaviour for the two \hkl<a>-type slip systems. Both prismatic and basal distributions are shifted to smaller event sizes. The largest events also occur at smaller $\Delta{}\bar{\tau}$ for the high-O alloy than for the quenched low-O material. These observations suggest that higher interstitial oxygen content promotes smoother slip intermittency, involving a larger number of smaller slip events.

For the aged low-O alloy, the introduction of \textalpha{}$_2$ shifts the CCDFs to the right, Fig.~\ref{fig:results-hedmstats}(c), indicating that \textalpha{}$_2$ ordered domains cause coarsening of slip intermittency. For the macroscopic strains of $\sim$1.5\% applied in our stress relaxation experiments, it appears that sufficient Al ordering remains to prevent substantial cross-slip and dislocation bowing within the slip bands, Fig.~\ref{fig:tem-dislocations}. For both of the \textalpha{}$_2$-containing materials, the HEDM CCDFs comprise a population of smaller stress drops, due to continued slip within existing bands that is gradually limited by back-stress from pile-ups at grain boundaries, with a raised incidence of large stress drops corresponding to the formation of entirely new slip bands as required by continued macroscopic strain.

In the high-O aged alloy, Fig.~\ref{fig:results-hedmstats}(d), the refined \textalpha{}$_2$ dispersion and consequent difficulty of setting up slip bands\textcolor{black}{, Fig.~\ref{fig:tem-dislocations},} resulted in even greater coarseness in this sense, with significant numbers of very large events compared with the other materials. There were also fewer stress drop events overall in the aged materials than in the quenched materials for equivalent macroscopic creep strain. In contrast with the smoothing effect of interstitial oxygen, the ordering of aluminium promotes deleterious sporadic slip, of the nature that may enhance the likelihood of dwell fatigue crack initiation.

\textcolor{black}{Regarding the combined or competing effects of oxygen and \textalpha{}$_2$ in the aged high-oxygen alloy, the post-deformation STEM images (Fig.~\ref{fig:tem-dislocations}) suggest that the effects of ordered domains in constraining dislocation motion outweigh the beneficial promotion of cross-slip by interstitial oxygen. We note further that oxygen segregates to the \textalpha{} phase upon precipitation of \textalpha{}$_2$ during ageing \cite{Dear2021}, so that an even greater \textalpha{} cross-slip tendency might have been expected for this specimen. The \textalpha{}$_2$ precipitates are fully coherent with the surrounding \textalpha{} matrix, so that particle shearing rather than dislocation bowing is the mechanism for dislocations to travel through the precipitate dispersion. As such, it appears that any enhanced cross-slip ability due to interstitial oxygen enrichment does not confer an advantage to dislocations trying to traverse an \textalpha{}$_2$-containing \textalpha{} grain. Therefore in the CCDFs, Fig.~\ref{fig:results-hedmstats}, the beneficial effect of O in raising the frequency of small pri\hkl<a> slip-associated load drops in Fig.~\ref{fig:results-hedmstats}(b) is suppressed by ageing, Fig.~\ref{fig:results-hedmstats}(d).}

\section{Discussion}

The above experimental results reveal factors influencing slip behaviour of \textalpha{}-titanium in the spatial and temporal domains. These behaviours underpin both the propensity to form slip bands and the impact of these bands on subsequent time-dependent deformation processes.

In the Ti--Al--O alloy system, the presence of either or both high interstitial oxygen content or crystallographic ordering of aluminium can cause slip band formation in the early stages of plasticity. Subsequent deformation predominantly takes place within these slip bands, but the strictness of this constraint depends on which factors underlie the initial slip band formation.

In the case of oxygen, dislocation motion is controlled by pinning and release by interstitial atoms. Short-range repulsion between O interstitials and dislocation cores means that cross-slip and movement of the O atom between interstices around the core are required to unpin the dislocation \cite{Ghazisaeidi2014,Yu2015}. The dependence of slip planarity on temperature, strain rate and oxygen content is due to the ability or otherwise of oxygen to be moved out of the glide plane of a travelling dislocation, via a shuffling mechanism \cite{Chong2020}. This cleared-out glide plane is then an easy route for further slip, evident in Fig.~\ref{fig:tem-dislocations}. There are also noticeable amounts of dislocation interaction within slip bands, and several dislocations extending between them. Chong and Zhang et al. \cite{Chong2021} explained their observations of similar behaviour in Ti--Al--O alloys at cryogenic temperatures as due to hardening within slip bands that promotes the emission of slip into relatively softer surrounding material, promoting spatial homogenisation of slip within a grain.

Slip band formation due to crystallographic ordering of aluminium is well known and occurs in materials with either short-range order (SRO) or long-range order in Ti$_3$Al \textalpha{}$_2$ precipitates. In both cases, dislocation motion is hindered due to formation of an antiphase boundary (APB, in \textalpha{}$_2$) or diffuse antiphase boundary (DAPB, in SRO) upon passage of the first dislocation. A second, trailing dislocation following the first can partially restore the ordering state, resulting in paired dislocation motion \cite{vandeWalle2002}. With continued slip, this results in progressive destruction of ordering, with other authors observing this after tensile deformation \cite{Williams2002} and during \textit{in situ} micropillar compression \cite{Zhang2019}. The requirement for paired motion also restricts the ability to cross-slip, limiting forest hardening. In contrast with the interstitial oxygen case, then, slip bands in \textalpha{}$_2$-strengthened material undergo irreversible softening, and there is no ability to extend slip into surrounding material to homogenise intragranular slip.

These distinct mechanisms result in the formation of inherently different types of slip band in \textalpha{}-titanium where interstitial O and ordering of Al are concerned, despite spatial heterogeneity appearing in both cases. The differences are reflected in the time intermittency of slip in the model materials studied here. Experimental work found that raised interstitial oxygen content promotes smoother time dependency of slip, while, \textalpha{}$_2$ causes coarser slip intermittency, with larger stress drop magnitudes and more unusually large events.

In disordered Ti--7Al--0.25O, we observed that slip proceeded by a larger frequency of smaller events than for Ti--7Al--0.05O to achieve the same macroscopic strain. Barkia et al. \cite{Barkia2017} tracked dislocation motion with \textit{in situ} TEM for two grades of CP Ti containing low and high amounts of oxygen comparable to those studied in the present work. They reported intermittent slip in both materials, with dislocations jumping forward by smaller distances at higher oxygen content, and noted that oxygen promotes continual cross-slip as dislocations change paths to avoid the oxygen atom obstacles. Our work shows that this behaviour also occurs in Ti--7Al, a model system for widely-used \textalpha{}+\textbeta{} structural alloys.

A mechanistic picture of the micromechanics emerges in which the evolving strength contrast between slip bands and surrounding matrix material is a key factor connecting the spatial and temporal heterogeneity of slip in Ti--Al--O alloys.

In the presence of raised interstitial oxygen content, dislocations must cross-slip and shuffle the oxygen atoms around to unpin themselves and glide to accommodate the applied strain. With the glide plane cleared of oxygen atoms, this region of the grain becomes the easiest route for subsequent slip, resulting in slip band formation. As more and more dislocations are added to the band, sufficient back-stress is eventually generated to drive the formation of new slip bands, or to promote dislocation motion in the unslipped material between bands. Since cross-slip is promoted by the high concentration of oxygen interstitials, more dislocation interactions can occur and contribute to beneficial forest hardening. This homogenises the spatial distribution of slip within the grain, reflected in the time domain as a series of many small events.

\textcolor{black}{Discrete dislocation plasticity modelling has also been conducted to investigate the relationship between obstacle spacing (i.e. spacing of oxygen interstitials) and the intermittency of thermally activated unpinning of dislocations. The model reproduces the experimentally observed differences in intermittency for the two disordered samples containing different oxygen concentrations, as well as the different behaviours of the basal and prismatic slip systems, to be reported in a later publication.}

With crystallographically ordered domains, meanwhile, the strength contrast between slip bands and surrounding unslipped material is much larger, and cross-slip is also restricted. The stresses required to trigger a new slip band to form are then significantly higher than in the oxygen case. The divergence of properties between slip bands and unslipped matrix makes it very difficult to engage the rest of the material in the grain in plasticity. Without a mechanism for spatial homogenisation, time intermittency becomes coarser due to the sudden formation of and activity within new slip bands once sufficient stresses are finally reached. This is even more extreme when the precipitate dispersion is refined, since the strength of the unslipped material is accordingly higher.

Slip band formation is inevitable to some degree for alloys with $\sim$\SI{6}{\wtpercent}~Al, which cannot be realistically industrially processed to achieve either zero crystallographic ordering of Al, or zero interstitial O content. Given this, and given the deleterious effect of intense and uncontrolled slip bands in promoting dwell fatigue crack initiation, the important consideration becomes the ability of the material to dynamically homogenise intragranular slip with continued deformation. Interstitial oxygen promotes this dynamic homogenisation, while crystallographic ordering of aluminium causes irretrievable divergence of the mechanical properties within and outside of slip bands in a grain. These results point towards alloying strategies for mitigation of runaway slip band intensification.

In summary, stress drops in individual grains during creep holds were observed in $\alpha$-Ti, following Beaudoin et al. using HEDM.  The effect of interstitial O and $\alpha_2$ precipitation were examined in Ti--7Al, which is a model alloy widely used to represent the $\alpha$ phase in commercial aerospace titanium alloys. The problem of cold dwell fatigue crack initiation is in view, where holds at relatively moderate stress can result in dramatic reductions in fatigue life in service in critical rotating parts in jet engines in certain situations. It is observed that O promotes slip homogeneity, with a higher frequency of smaller stress drops being observed, whereas $\alpha_2$ precipitation results in fewer, larger events. As expected, basal slip is more common and gives larger slip events. These observations are in concordance with the emerging picture that moving \hkl<a> dislocations can clear oxygen from their slip planes, while $\alpha_2$ domains must be sheared by pairs of dislocations, restricting cross-slip. This suggests stress redistribution from soft basal-activated units onto adjacent hard units would be higher, potentially explaining independent observations \cite{AFRL1757} that dwell facet formation is often observed to be associated with basal slip.

\section{Methods}\label{methods}\small

\textbf{Experimental methods.} Alloys of nominal composition Ti--7Al--0.05O and Ti--7Al--0.25O (\si{\wtpercent}) were melted from pure elements and TiO powder using an Arcast200 \SI{27}{\kilo\watt} low pressure argon arc melter. Ingots were rolled into square bars, with \textbeta{} breakdown and \textalpha{} rolling stages followed by recrystallisation in the \textalpha{} field, as described in \cite{Dear2021}. The bars were quenched in ice water from the recrystallisation temperature, producing a crystallographically disordered equiaxed \textalpha{} microstructure. Sections of each bar were then encapsulated in an argon atmosphere in quartz ampoules, and aged at \SI{550}{\celsius} for \SI{49}{\day} followed by furnace cooling. This produced a fine, homogeneous dispersion of \textalpha{}$_2$ precipitates in the aged specimens. Under these ageing conditions, the volume fractions of \textalpha{}$_2$ are similar in the two alloys \cite{Dear2021}.

\begin{figure}[t!]
   \centering
   \includegraphics[width=\columnwidth]{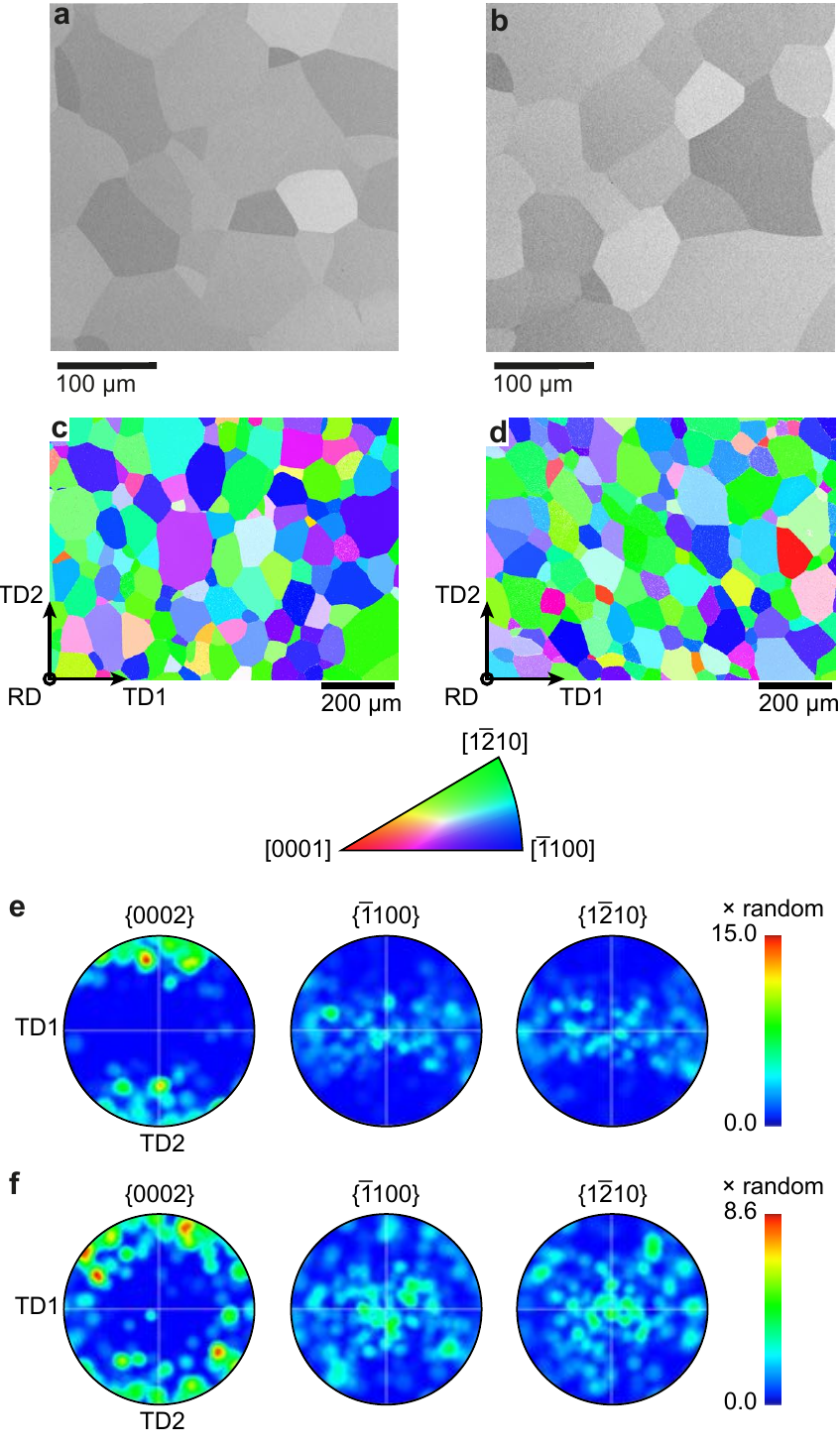}
   \caption{Backscatter electron images of \textbf{a} Ti--7Al--0.05O and \textbf{b} Ti--7Al--0.25O (wt.\%) alloys, and \textbf{c}, \textbf{d} EBSD data displaying their respective textures. For EBSD, inverse pole figure maps are with respect to the rolling direction of the bar. The rolling direction was parallel to tensile axis in tensile, stress relaxation and creep tests in this work. Note the paucity of grains with \textbf{c}-axis oriented close to the loading direction, meaning that `hard' grain orientations were only rarely captured in the analysis volume for HEDM.}
   \label{fig:supplementary-sem}
\end{figure}

Microstructures were characterised using backscatter electron imaging and electron backscatter diffraction on a Zeiss Sigma~300 FEG-SEM, using \SI{8}{\kilo\volt} and \SI{20}{\kilo\volt} accelerating voltages respectively, showing an equiaxed single-phase \textalpha{} microstructure with typical bar rolling texture, Fig.~\ref{fig:supplementary-sem}. EBSD was performed using a Bruker eFlash detector. Dislocation configurations and, where present, \textalpha{}$_2$ dispersions were observed with dark-field imaging using a JEOL 2100F TEM operated at \SI{200}{\kilo\volt}. Microscopy specimens were prepared by electropolishing with a 3\% perchloric acid solution at \SI{-35}{\celsius} and \SI{20}{\volt}.

For initial evaluation of mechanical properties, tensile testing was conducted on specimens with a 1.5$\times$1.5$\times$\SI{19}{\milli\metre} gauge geometry, using an Instron 5967 load frame with a \SI{30}{\kilo\newton} load cell and an Instron 2620 extensometer, at a strain rate of \SI{E-3}{\per\second}. Tests were paused momentarily upon reaching 10\% strain for removal of the extensometer prior to sample fracture where possible. Stress relaxation tests were conducted on the same equipment, using a 3$\times$3$\times$\SI{19}{\milli\metre} gauge geometry that permitted direct extraction of electropolished TEM specimens from the gauge. In these tests, specimens were loaded to a chosen strain at a strain rate of \SI{E-4}{\per\second}, before holding for \textcolor{black}{\SI{600}{\second}} in position control to allow relaxation of macroscopic tensile stresses.

\textit{In situ} creep measurements were performed with far-field high energy X-ray diffraction microscopy (ff-HEDM) scans on the FAST beamline at the Cornell High Energy Synchrotron Source (CHESS). The RAMS2 load frame was used, capable of applying tensile loads and rotating the entire specimen/load frame assembly in the beam \cite{Shade2015}. Specimens with a 1$\times$\SI{1}{\milli\metre} gauge section were fabricated by precision wire electro-discharge machining provided through Die Technology Inc. Macroscopic strain was tracked using digital image correlation, using the as-machined surface as a speckle pattern, and macroscopic stress was measured using a load cell. Specimens were scanned at zero load, before stepping up to \SI{450}{\mega\pascal} and incrementally approaching the desired creep stress using \SI{10}{\mega\pascal} steps. For the ff-HEDM measurements, specimens were illuminated with a \SI{1}{\milli\metre} beam height, capturing a \SI{1}{\milli\metre\cubed} volume. Far field patterns were recorded using a beam energy of \SI{41.991}{\kilo\electronvolt} and two Dexela 2923 NDT panel detectors located \SI{878}{\milli\metre} from the specimen. Far field patterns were recorded at \SI{0.25}{\degree} increments to cover a \SI{360}{\degree} rotation for each scan, resulting in 1,440 frames per scan. Each scan took approximately \SI{6}{\minute}, taken without unloading the specimens or interrupting the creep experiments. The resulting ff-HEDM data were reduced using the \texttt{hexrd} package \cite{hexrd}, and subsequent analysis performed in Python.

%\section*{References}
%\bibliography{sample.bib}

\begin{thebibliography}{40}
\expandafter\ifx\csname natexlab\endcsname\relax\def\natexlab#1{#1}\fi
\providecommand{\bibinfo}[2]{#2}
\ifx\xfnm\relax \def\xfnm[#1]{\unskip,\space#1}\fi
%Type = Article
\bibitem[{Carmen-Miguel et~al.(2001)Carmen-Miguel, Vespignani, Zapperi, Weiss,
  and Grasso}]{Miguel2001}
\bibinfo{author}{M.~Carmen-Miguel}, \bibinfo{author}{A.~Vespignani},
  \bibinfo{author}{S.~Zapperi}, \bibinfo{author}{J.~Weiss},
  \bibinfo{author}{J.-R. Grasso},
\newblock \bibinfo{title}{Intermittent dislocation flow in viscoplastic
  deformation},
\newblock \bibinfo{journal}{Nature} \bibinfo{volume}{410}
  (\bibinfo{year}{2001}) \bibinfo{pages}{667--671}.
%Type = Article
\bibitem[{Weiss et~al.(2007)Weiss, Richeton, Louchet, Chmelik, Dobron,
  Entemeyer, Lebyodkin, Lebedkina, Fressengeas, and McDonald}]{Weiss2007}
\bibinfo{author}{J.~Weiss}, \bibinfo{author}{T.~Richeton},
  \bibinfo{author}{F.~Louchet}, \bibinfo{author}{F.~Chmelik},
  \bibinfo{author}{P.~Dobron}, \bibinfo{author}{D.~Entemeyer},
  \bibinfo{author}{M.~Lebyodkin}, \bibinfo{author}{T.~Lebedkina},
  \bibinfo{author}{C.~Fressengeas}, \bibinfo{author}{R.~J. McDonald},
\newblock \bibinfo{title}{Evidence for universal intermittent crystal
  plasticity from acoustic emission and high-resolution extensometry
  experiments},
\newblock \bibinfo{journal}{Phys. Rev. B} \bibinfo{volume}{76}
  (\bibinfo{year}{2007}) \bibinfo{pages}{224110}.
%Type = Article
\bibitem[{Brown(2012)}]{Brown2012}
\bibinfo{author}{L.~M. Brown},
\newblock \bibinfo{title}{Constant intermittent flow of dislocations: {C}entral
  problems in plasticity},
\newblock \bibinfo{journal}{Mater. Sci. Tech.} \bibinfo{volume}{28}
  (\bibinfo{year}{2012}) \bibinfo{pages}{1209--1232}.
%Type = Article
\bibitem[{Sparks and Maa{\ss}(2018)}]{Sparks2018}
\bibinfo{author}{G.~Sparks}, \bibinfo{author}{R.~Maa{\ss}},
\newblock \bibinfo{title}{{Shapes and velocity relaxation of dislocation
  avalanches in Au and Nb microcrystals}},
\newblock \bibinfo{journal}{Acta Mater.} \bibinfo{volume}{152}
  (\bibinfo{year}{2018}) \bibinfo{pages}{86--95}.
%Type = Book
\bibitem[{Christensen(1982)}]{Christensen1982}
\bibinfo{author}{R.~M. Christensen}, \bibinfo{title}{Theory of viscoelasticity:
  {A}n introduction}, \bibinfo{publisher}{Academic Press},
  \bibinfo{edition}{second} edition, \bibinfo{year}{1982}.
%Type = Article
\bibitem[{Zheng et~al.(2016)Zheng, Balint, and Dunne}]{Zheng2016}
\bibinfo{author}{Z.~Zheng}, \bibinfo{author}{D.~S. Balint},
  \bibinfo{author}{F.~P.~E. Dunne},
\newblock \bibinfo{title}{Dwell fatigue in two {T}i alloys: {A}n integrated
  crystal plasticity and discrete dislocation study},
\newblock \bibinfo{journal}{J. Mech. Phys. Solids} \bibinfo{volume}{96}
  (\bibinfo{year}{2016}) \bibinfo{pages}{411--427}.
%Type = Article
\bibitem[{Bridier et~al.(2008)Bridier, Villechaise, and Mendez}]{Bridier2008}
\bibinfo{author}{F.~Bridier}, \bibinfo{author}{P.~Villechaise},
  \bibinfo{author}{J.~Mendez},
\newblock \bibinfo{title}{Slip and fatigue crack formation processes in an
  \textalpha{}/\textbeta{} titanium alloy in relation to crystallographic
  texture on different scales},
\newblock \bibinfo{journal}{Acta Mater.} \bibinfo{volume}{56}
  (\bibinfo{year}{2008}) \bibinfo{pages}{3951--3962}.
%Type = Article
\bibitem[{Neal(1988)}]{Neal1988}
\bibinfo{author}{D.~F. Neal},
\newblock \bibinfo{title}{Creep fatigue interactions in titanium alloys},
\newblock \bibinfo{journal}{Proceedings of the Sixth World Conference on
  Titanium}  (\bibinfo{year}{1988}) \bibinfo{pages}{175--180}.
%Type = Article
\bibitem[{Dunne et~al.(2007)Dunne, Rugg, and Walker}]{Dunne2007}
\bibinfo{author}{F.~P.~E. Dunne}, \bibinfo{author}{D.~Rugg},
  \bibinfo{author}{A.~Walker},
\newblock \bibinfo{title}{Lengthscale-dependent, elastically anisotropic,
  physically-based hcp crystal plasticity: {A}pplication to cold-dwell fatigue
  in {T}i alloys},
\newblock \bibinfo{journal}{Int. J. Plasticity} \bibinfo{volume}{23}
  (\bibinfo{year}{2007}) \bibinfo{pages}{1061--1083}.
%Type = Article
\bibitem[{Mills et~al.(2017)Mills, Ghosh, Rokhlin, Brandes, Pilchak, and
  Williams}]{AFRL1757}
\bibinfo{author}{M.~J. Mills}, \bibinfo{author}{S.~Ghosh},
  \bibinfo{author}{S.~Rokhlin}, \bibinfo{author}{M.~C. Brandes},
  \bibinfo{author}{A.~L. Pilchak}, \bibinfo{author}{J.~C. Williams},
\newblock \bibinfo{title}{{The Evaluation of Cold Dwell Fatigue in Ti-6242}},
\newblock \bibinfo{journal}{Final Report, DOT/FAA/TC-17/57, Federal Aviation
  Administration}  (\bibinfo{year}{2017}).
%Type = Article
\bibitem[{Neeraj et~al.(2000)Neeraj, Hou, Daehn, and Mills}]{Neeraj2000}
\bibinfo{author}{T.~Neeraj}, \bibinfo{author}{D.-H. Hou},
  \bibinfo{author}{G.~S. Daehn}, \bibinfo{author}{M.~J. Mills},
\newblock \bibinfo{title}{Phenomenological and microstructural analysis of room
  temperature creep in titanium alloys},
\newblock \bibinfo{journal}{Acta Mater.} \bibinfo{volume}{48}
  (\bibinfo{year}{2000}) \bibinfo{pages}{1225--1238}.
%Type = Article
\bibitem[{Bache(2003)}]{Bache2003}
\bibinfo{author}{M.~R. Bache},
\newblock \bibinfo{title}{A review of dwell sensitive fatigue in titanium
  alloys: {T}he role of microstructure, texture and operating conditions},
\newblock \bibinfo{journal}{Int. J. Fatigue} \bibinfo{volume}{25}
  (\bibinfo{year}{2003}) \bibinfo{pages}{1079--1087}.
%Type = Article
\bibitem[{Yu et~al.(2015)Yu, Qi, Tsuru, Traylor, Rugg, {J. W. Morris Jr.},
  Asta, Chrzan, and Minor}]{Yu2015}
\bibinfo{author}{Q.~Yu}, \bibinfo{author}{L.~Qi}, \bibinfo{author}{T.~Tsuru},
  \bibinfo{author}{R.~Traylor}, \bibinfo{author}{D.~Rugg}, \bibinfo{author}{{J.
  W. Morris Jr.}}, \bibinfo{author}{M.~D. Asta}, \bibinfo{author}{D.~C.
  Chrzan}, \bibinfo{author}{A.~M. Minor},
\newblock \bibinfo{title}{{Origin of dramatic oxygen solute strengthening
  effect in titanium}},
\newblock \bibinfo{journal}{{Science}} \bibinfo{volume}{{347}}
  (\bibinfo{year}{{2015}}) \bibinfo{pages}{{635--639}}.
%Type = Article
\bibitem[{Barkia et~al.(2017)Barkia, Couzini\'{e}, Lartigue-Korinek, Guillot,
  and Doquet}]{Barkia2017}
\bibinfo{author}{B.~Barkia}, \bibinfo{author}{J.~P. Couzini\'{e}},
  \bibinfo{author}{S.~Lartigue-Korinek}, \bibinfo{author}{I.~Guillot},
  \bibinfo{author}{V.~Doquet},
\newblock \bibinfo{title}{{\textit{In situ} TEM observations of dislocation
  dynamics in \textalpha{} titanium: Effect of the oxygen content}},
\newblock \bibinfo{journal}{Mat. Sci. Eng. A--Struct.} \bibinfo{volume}{703}
  (\bibinfo{year}{2017}) \bibinfo{pages}{331--339}.
%Type = Article
\bibitem[{Chong et~al.(2020)Chong, Poschmann, Zhang, Zhao, Hooshmand,
  Rothchild, Olmsted, {J. W. Morris Jr.}, Chrzan, Asta, and Minor}]{Chong2020}
\bibinfo{author}{Y.~Chong}, \bibinfo{author}{M.~Poschmann},
  \bibinfo{author}{R.~Zhang}, \bibinfo{author}{S.~Zhao}, \bibinfo{author}{M.~S.
  Hooshmand}, \bibinfo{author}{E.~Rothchild}, \bibinfo{author}{D.~L. Olmsted},
  \bibinfo{author}{{J. W. Morris Jr.}}, \bibinfo{author}{D.~C. Chrzan},
  \bibinfo{author}{M.~Asta}, \bibinfo{author}{A.~M. Minor},
\newblock \bibinfo{title}{Mechanistic basis of oxygen sensitivity in titanium},
\newblock \bibinfo{journal}{Sci. Adv.} \bibinfo{volume}{6}
  (\bibinfo{year}{2020}) \bibinfo{pages}{eabc4060}.
%Type = Article
\bibitem[{Bernier et~al.(2011)Bernier, Barton, Lienert, and
  Miller}]{Bernier2011}
\bibinfo{author}{J.~V. Bernier}, \bibinfo{author}{N.~R. Barton},
  \bibinfo{author}{U.~Lienert}, \bibinfo{author}{M.~P. Miller},
\newblock \bibinfo{title}{Far-field high-energy diffraction microscopy: {A}
  tool for intergranular orientation and strain analysis},
\newblock \bibinfo{journal}{J Strain Anal. Eng.} \bibinfo{volume}{46}
  (\bibinfo{year}{2011}) \bibinfo{pages}{527--547}.
%Type = Article
\bibitem[{Pagan et~al.(2017)Pagan, Shade, Barton, Park, Kenesei, Menasche, and
  Bernier}]{Pagan2017}
\bibinfo{author}{D.~C. Pagan}, \bibinfo{author}{P.~A. Shade},
  \bibinfo{author}{N.~R. Barton}, \bibinfo{author}{J.-S. Park},
  \bibinfo{author}{P.~Kenesei}, \bibinfo{author}{D.~B. Menasche},
  \bibinfo{author}{J.~V. Bernier},
\newblock \bibinfo{title}{{Modeling slip system strength evolution in Ti--7Al
  informed by \textit{in-situ} grain stress measurements}},
\newblock \bibinfo{journal}{Acta Mater.} \bibinfo{volume}{{128}}
  (\bibinfo{year}{{2017}}) \bibinfo{pages}{406--417}.
%Type = Article
\bibitem[{Xu et~al.(2020)Xu, Joseph, Karamched, Fox, Rugg, Dunne, and
  Dye}]{XuJoseph2020}
\bibinfo{author}{Y.~Xu}, \bibinfo{author}{S.~Joseph},
  \bibinfo{author}{P.~Karamched}, \bibinfo{author}{K.~Fox},
  \bibinfo{author}{D.~Rugg}, \bibinfo{author}{F.~P.~E. Dunne},
  \bibinfo{author}{D.~Dye},
\newblock \bibinfo{title}{Predicting dwell fatigue life in titanium alloys
  using modelling and experiment},
\newblock \bibinfo{journal}{Nat. Commun.} \bibinfo{volume}{11}
  (\bibinfo{year}{2020}) \bibinfo{pages}{5868}.
%Type = Article
\bibitem[{Isp\'{a}novity~et~al.(2022)Isp\'{a}novity, Ugi, P\'{e}terffy, Knapek, Kal\'{a}cska, T\"{u}zes, Dankh\'{a}zi, M\'{a}this, Chmel\'{i}k and Groma}]{Ispanovity2022}
\bibinfo{author}{P\'{e}ter Dus\'{a}n Isp\'{a}novity}, \bibinfo{author}{D\'{a}vid Ugi},
  \bibinfo{author}{G\'{a}bor P\'{e}terffy}, 
  \bibinfo{author}{Michal Knapek},
  \bibinfo{author}{Szilvia Kal\'{a}cska}, \bibinfo{author}{D\'{a}niel T\"{u}zes},
  \bibinfo{author}{Zolt\'{a}n Dankh\'{a}zi},
  \bibinfo{author}{Kristi\'{a}n M\'{a}this},  \bibinfo{author}{Franti\v{s}ek Chmel\'{i}k},  \bibinfo{author}{Istv\'{a}n Groma},
\newblock \bibinfo{title}{Dislocation avalanches are like earthquakes on the micron scale},
\newblock \bibinfo{journal}{Nat. Commun.} \bibinfo{volume}{13}
  (\bibinfo{year}{2022}) \bibinfo{pages}{1975}.
%Type = Article
\bibitem[{Beaudoin et~al.(2017)Beaudoin, Shade, Schuren, Turner, Woodward,
  Bernier, Li, Dimiduk, Kenesei, and Park}]{Beaudoin2017}
\bibinfo{author}{A.~J. Beaudoin}, \bibinfo{author}{P.~A. Shade},
  \bibinfo{author}{J.~C. Schuren}, \bibinfo{author}{T.~J. Turner},
  \bibinfo{author}{C.~Woodward}, \bibinfo{author}{J.~V. Bernier},
  \bibinfo{author}{S.~F. Li}, \bibinfo{author}{D.~M. Dimiduk},
  \bibinfo{author}{P.~Kenesei}, \bibinfo{author}{J.-S. Park},
\newblock \bibinfo{title}{Bright {X}-rays reveal shifting deformation states
  and effects of the microstructure on the plastic deformation of crystalline
  materials},
\newblock \bibinfo{journal}{Phys. Rev. B} \bibinfo{volume}{96}
  (\bibinfo{year}{2017}) \bibinfo{pages}{174116}.
%Type = Article
%\bibitem[{van~der Giessen and Needleman(1995)}]{vanderGiessen1995}
%\bibinfo{author}{E.~van~der Giessen}, \bibinfo{author}{A.~Needleman},
%\newblock \bibinfo{title}{{Discrete Dislocation Plasticity - a Simple Planar
%  Model}},
%\newblock \bibinfo{journal}{Modelling Simul. Mater. Sci. Eng.}
%  \bibinfo{volume}{3} (\bibinfo{year}{1995}) \bibinfo{pages}{689--735}.
% %Type = Article
%\bibitem[{Zheng et al.(2016)}]{Zheng2016b}
%\bibinfo{author}{Z. Zheng}, \bibinfo{author}{D. Balint}, \bibinfo{author}{F.P.E. Dunne},
%\newblock \bibinfo{title}{{Rate sensitivity in discrete dislocation plasticity in hexagonal close-packed crystals}},
%\newblock \bibinfo{journal}{Acta Mater.}
%  \bibinfo{volume}{107} (\bibinfo{year}{2016}) \bibinfo{pages}{17--26}.
%Type = Article
%\bibitem[{Zhang et~al.(2016)Zhang, Jun, Britton, and Dunne}]{Zhang2016}
%\bibinfo{author}{Z.~Zhang}, \bibinfo{author}{T.-S. Jun}, \bibinfo{author}{T.~B.
%  Britton}, \bibinfo{author}{F.~P.~E. Dunne},
%\newblock \bibinfo{title}{Intrinsic anisotropy of strain rate sensitivity in
%  single crystal alpha titanium},
%\newblock \bibinfo{journal}{Acta Mater.} \bibinfo{volume}{118}
%  (\bibinfo{year}{2016}) \bibinfo{pages}{317--330}.
%Type = Article
\bibitem[{Dear et~al.(2021)Dear, Kontis, Gault, Ilavsky, Rugg, and
  Dye}]{Dear2021}
\bibinfo{author}{F.~F. Dear}, \bibinfo{author}{P.~Kontis},
  \bibinfo{author}{B.~Gault}, \bibinfo{author}{J.~Ilavsky},
  \bibinfo{author}{D.~Rugg}, \bibinfo{author}{D.~Dye},
\newblock \bibinfo{title}{{Mechanisms of Ti$_3$Al precipitation in hcp
  \textalpha{}-Ti}},
\newblock \bibinfo{journal}{Acta Mater.} \bibinfo{volume}{212}
  (\bibinfo{year}{2021}) \bibinfo{pages}{116811}.
%Type = Article
\bibitem[{Lim et~al.(1976)Lim, McMahon, Pope, Williams}]{Lim1976}
\bibinfo{author}{J.~Y. Lim}, \bibinfo{author}{C.~J. McMahon, Jr.},
  \bibinfo{author}{D.~P. Pope}, \bibinfo{author}{J.~C. Williams},
  \bibinfo{title}{{The Effect of Oxygen on the Structure and Mechanical Behavior of Aged Ti--8 Wt Pct Al}}, \bibinfo{year}{1976}.
%Type = Article
\bibitem[{Williams et~al.(1972)Williams, Sommer, and Tung}]{Williams1972}
\bibinfo{author}{J.~C. Williams}, \bibinfo{author}{A.~W. Sommer},
  \bibinfo{author}{P.~P. Tung},
\newblock \bibinfo{title}{The influence of oxygen concentration on the internal
  stress and dislocation arrangements in \textalpha{} titanium},
\newblock \bibinfo{journal}{Metall. Mater. Trans. B} \bibinfo{volume}{3}
  (\bibinfo{year}{1972}) \bibinfo{pages}{2979--2984}.
%Type = Article
\bibitem[{Neeraj and Mills(2001)}]{Neeraj2001}
\bibinfo{author}{T.~Neeraj}, \bibinfo{author}{M.~J. Mills},
\newblock \bibinfo{title}{{Short-range order (SRO) and its effect on the
  primary creep behavior of a Ti--6wt.\% Al alloy}},
\newblock \bibinfo{journal}{Mat. Sci. Eng. A--Struct.} \bibinfo{volume}{319}
  (\bibinfo{year}{2001}) \bibinfo{pages}{415--419}.

%Type = Article
\bibitem[{Zhang et~al.(2019)Zhang, Zhao, Ophus, Deng, Vachhani, Ozdo, Traylor,
  Bustillo, {J. W. Morris Jr.}, Chrzan, Asta, and Minor}]{Zhang2019}
\bibinfo{author}{R.~Zhang}, \bibinfo{author}{S.~Zhao},
  \bibinfo{author}{C.~Ophus}, \bibinfo{author}{Y.~Deng}, \bibinfo{author}{S.~J.
  Vachhani}, \bibinfo{author}{B.~Ozdo}, \bibinfo{author}{R.~Traylor},
  \bibinfo{author}{K.~C. Bustillo}, \bibinfo{author}{{J. W. Morris Jr.}},
  \bibinfo{author}{D.~C. Chrzan}, \bibinfo{author}{M.~Asta},
  \bibinfo{author}{A.~M. Minor},
\newblock \bibinfo{title}{{Direct imaging of short-range order and its impact
  on deformation in Ti--6Al}},
\newblock \bibinfo{journal}{Sci. Adv.} \bibinfo{volume}{5}
  (\bibinfo{year}{2019}) \bibinfo{pages}{eaax2799}.
 %Type = Article
%\bibitem[{Gibbs(1969)}]{Gibbs1969}
%\bibinfo{author}{G. B. Gibbs},
%\newblock \bibinfo{title}{{Thermodynamic analysis of dislocation glide controlled by dispersed local obstacles.}},
%\newblock \bibinfo{journal}{Mat. Sci. Eng.} \bibinfo{volume}{4(6)}
%  (\bibinfo{year}{1969}) \bibinfo{pages}{313--328}.
%
%Type = Article
%\bibitem[{Weiss and Marsan(2003)}]{Weiss2003}
%\bibinfo{author}{J.~Weiss}, \bibinfo{author}{D.~Marsan},
%\newblock \bibinfo{title}{{Three-Dimensional Mapping of Dislocation Avalanches:
%  Clustering and Space/Time Coupling}},
%\newblock \bibinfo{journal}{Science} \bibinfo{volume}{299}
%  (\bibinfo{year}{2003}) \bibinfo{pages}{89--92}.
%Type = Article
%\bibitem[{Richeton et~al.(2005)Richeton, Weiss, and Louchet}]{Richeton2005}
%\bibinfo{author}{T.~Richeton}, \bibinfo{author}{J.~Weiss},
%  \bibinfo{author}{F.~Louchet},
%\newblock \bibinfo{title}{Breakdown of avalanche critical behaviour in
%  polycrystalline plasticity},
%\newblock \bibinfo{journal}{Nat. Mater.} \bibinfo{volume}{4}
%  (\bibinfo{year}{2005}) \bibinfo{pages}{465--469}.
%Type = Article
%\bibitem[{Crosby et~al.(2015)Crosby, Po, Erel, and Ghoniem}]{Crosby2015}
%\bibinfo{author}{T.~Crosby}, \bibinfo{author}{G.~Po},
%  \bibinfo{author}{C.~Erel}, \bibinfo{author}{N.~Ghoniem},
%\newblock \bibinfo{title}{The origin of strain avalanches in sub-micron
%  plasticity of fcc metals},
%\newblock \bibinfo{journal}{Acta Mater.} \bibinfo{volume}{89}
%  (\bibinfo{year}{2015}) \bibinfo{pages}{123--132}.
%Type = Article
%\bibitem[{Uchic et~al.(2004)Uchic, Dimiduk, Florando, and Nix}]{Uchic2004}
%\bibinfo{author}{M.~D. Uchic}, \bibinfo{author}{D.~M. Dimiduk},
%  \bibinfo{author}{J.~N. Florando}, \bibinfo{author}{W.~D. Nix},
%\newblock \bibinfo{title}{Sample dimensions influence strength and crystal
%  plasticity},
%\newblock \bibinfo{journal}{Science} \bibinfo{volume}{305}
%  (\bibinfo{year}{2004}) \bibinfo{pages}{986--989}.
%Type = Article
%\bibitem[{Dimiduk et~al.(2006)Dimiduk, Woodward, Lesar, and
%  Uchic}]{Dimiduk2006}
%\bibinfo{author}{D.~M. Dimiduk}, \bibinfo{author}{C.~Woodward},
%  \bibinfo{author}{R.~Lesar}, \bibinfo{author}{M.~D. Uchic},
%\newblock \bibinfo{title}{Scale-free intermittent flow in crystal plasticity},
%\newblock \bibinfo{journal}{Science} \bibinfo{volume}{312}
%  (\bibinfo{year}{2006}) \bibinfo{pages}{1188--1190}.
%Type = Article
\bibitem[{Warwick et~al.(2014)}]{Warwick2012}
\bibinfo{author}{J. L. W. Warwick}, \bibinfo{author}{J. Coakley}, \bibinfo{author}{S. L. Raghunathan}, \bibinfo{author}{R. J. Talling}, \bibinfo{author}{D. Dye},
\newblock \bibinfo{title}{{Effect of texture on load partitioning in Ti--6Al--4V}},
\newblock \bibinfo{journal}{Acta Mater.} \bibinfo{volume}{60}
  (\bibinfo{year}{2012}) \bibinfo{pages}{4117--4127}.
%Type = Article
\bibitem[{Ghazisaeidi and Trinkle(2014)}]{Ghazisaeidi2014}
\bibinfo{author}{M.~Ghazisaeidi}, \bibinfo{author}{D.~R. Trinkle},
\newblock \bibinfo{title}{Interaction of oxygen interstitials with lattice
  faults in {T}i},
\newblock \bibinfo{journal}{Acta Mater.} \bibinfo{volume}{76}
  (\bibinfo{year}{2014}) \bibinfo{pages}{82--86}.
%Type = Article
\bibitem[{Chong et~al.(2021)Chong, Zhang, Hooshmand, Zhao, Chrzan, Asta, {J. W.
  Morris Jr.}, and Minor}]{Chong2021}
\bibinfo{author}{Y.~Chong}, \bibinfo{author}{R.~Zhang}, \bibinfo{author}{M.~S.
  Hooshmand}, \bibinfo{author}{S.~Zhao}, \bibinfo{author}{D.~C. Chrzan},
  \bibinfo{author}{M.~Asta}, \bibinfo{author}{{J. W. Morris Jr.}},
  \bibinfo{author}{A.~M. Minor},
\newblock \bibinfo{title}{{Elimination of oxygen sensitivity in
  \textalpha{}-titanium by substitutional alloying with Al}},
\newblock \bibinfo{journal}{Nat. Commun.} \bibinfo{volume}{12}
  (\bibinfo{year}{2021}) \bibinfo{pages}{6158}.
%Type = Article
\bibitem[{van~de Walle and Asta(2002)}]{vandeWalle2002}
\bibinfo{author}{A.~van~de Walle}, \bibinfo{author}{M.~Asta},
\newblock \bibinfo{title}{{First-Principles Investigation of Perfect and
  Diffuse Antiphase Boundaries in HCP-Based Ti-Al Alloys}},
\newblock \bibinfo{journal}{Metall. Mater. Trans. A} \bibinfo{volume}{33A}
  (\bibinfo{year}{2002}) \bibinfo{pages}{735--741}.
%Type = Article
\bibitem[{Williams et~al.(2002)Williams, Baggerly, and Paton}]{Williams2002}
\bibinfo{author}{J.~C. Williams}, \bibinfo{author}{R.~G. Baggerly},
  \bibinfo{author}{N.~E. Paton},
\newblock \bibinfo{title}{Deformation behavior of {HCP Ti-Al} alloy single
  crystals},
\newblock \bibinfo{journal}{Metall. Mater. Trans. A} \bibinfo{volume}{33}
  (\bibinfo{year}{2002}) \bibinfo{pages}{837--850}.
%Type = Article
\bibitem[{Shade et~al.(2015)Shade, Blank, Schuren, Turner, Kenesei, Goetze,
  Suter, Bernier, Li, Lind, Lienert, and Almer}]{Shade2015}
\bibinfo{author}{P.~A. Shade}, \bibinfo{author}{B.~Blank},
  \bibinfo{author}{J.~C. Schuren}, \bibinfo{author}{T.~J. Turner},
  \bibinfo{author}{P.~Kenesei}, \bibinfo{author}{K.~Goetze},
  \bibinfo{author}{R.~M. Suter}, \bibinfo{author}{J.~V. Bernier},
  \bibinfo{author}{S.~F. Li}, \bibinfo{author}{J.~Lind},
  \bibinfo{author}{U.~Lienert}, \bibinfo{author}{J.~Almer},
\newblock \bibinfo{title}{{A rotational and axial motion system load frame
  insert for \textit{in situ} high energy X-ray studies}},
\newblock \bibinfo{journal}{Rev. Sci. Instrum.} \bibinfo{volume}{86}
  (\bibinfo{year}{2015}) \bibinfo{pages}{093902}.
%Type = Misc
\bibitem[{Barton et~al.(2021)Barton, Saransh, Bernier, Stowell, Harris, Major,
  Avery, Tourtellotte, Boyce, and Administration}]{hexrd}
\bibinfo{author}{N.~R. Barton}, \bibinfo{author}{F.~Saransh},
  \bibinfo{author}{J.~V. Bernier}, \bibinfo{author}{M.~L. Stowell},
  \bibinfo{author}{C.~Harris}, \bibinfo{author}{B.~Major},
  \bibinfo{author}{P.~Avery}, \bibinfo{author}{J.~Tourtellotte},
  \bibinfo{author}{D.~Boyce}, \bibinfo{author}{U.~N. N.~S. Administration},
  \bibinfo{title}{{Highly Extensible X-ray Diffraction Toolkit, Version
  0.8.1}}, \bibinfo{year}{2021}.

\end{thebibliography}

\vspace{0.5cm}
\noindent\textbf{Acknowledgements}
FFW was funded by Rolls-Royce plc and by the EPSRC Centre for Doctoral Training in the Advanced Characterisation of Materials (EP/L015277/1). FFW wishes to thank Dr Thomas J. Kwok for his assistance in preparing the alloys. Much of this work was initiated and funded through the Hexmat EPSRC program grant EP/K034332/1. DD also acknowledges provision of a Royal Society Industry Fellowship. YX acknowledges the financial support by the Engineering and Physical Sciences Research Council for funding through the grant EP/R018863/1. JVB acknowledges this work was performed under the auspices of the U.S. Department of Energy by Lawrence Livermore National Laboratory under Contract DE-AC52-07NA27344, and LDRD grant \#20-ERD-044 (LLNL-JRNL-835496). This work is based upon research conducted at the Center for High Energy X-ray Sciences (CHEXS) which is supported by the National Science Foundation under award DMR-1829070.

\noindent\textbf{Statement on Competing Interests.} One of us (D. Rugg) was employed by a jet engine manufacturer, Rolls-Royce plc, which also part-funded (with EPSRC) Dr Worsnop's PhD, which resulted in this work. Prof. Dye is a Royal Society funded Industry Fellow, with strong interaction with Rolls-Royce plc. The company has no direct stake in this work, nor has it been authorised or endorsed for publication by them.

\noindent\textbf{Author Contributions.} FFW processed the alloys and did the microscopy and mechanical experiments. FFW, JVB, DCP, REL, TPM and DD did the HEDM experiment. FFW, REL, DCP and JVB did the HEDM data reduction and analysis, using tools and techniques developed by REL, DCP and JVB. YX contributed to discussion and interpretation of the phenomena observed. DR, DD and FFW conceived the study. FFW and DD drafted the manuscript, which was edited by all the authors.

\noindent\textbf{Materials and Correspondence.}  Requests for material or data can be addressed to david.dye@imperial.ac.uk. 

\end{document}